\documentclass[final,10pt,times,twocolumn]{elsarticle}
\usepackage[%showframe,%Uncomment any one of the following lines to test 
margin=1.4cm,
]{geometry}
\usepackage{amsmath}
\usepackage{amsfonts}
\usepackage{amssymb}
\usepackage{graphicx}
\usepackage{lmodern}
\usepackage{hyperref}
\usepackage{dcolumn}% Align table columns on decimal point
\usepackage{bm}% bold math
\usepackage[mathlines]{lineno}% Enable numbering of text and display math
\usepackage{float}
\usepackage[T1]{fontenc}
\usepackage{subfigure}
\usepackage{natbib}
\usepackage{breqn}
\allowdisplaybreaks

\journal{Physics Letters B}

\usepackage{numcompress}%\bibliographystyle{model6-num-names}

\DeclareUnicodeCharacter{2212}{\ensuremath{-}}

\begin{document}
\newcolumntype{P}[1]{>{\centering\arraybackslash}p{#1}}
\title{\textbf{The Seesaw Evaded Modular Dirac Framework}}
%\tnotetext[mytitlenote]{Fully documented %templates are available in the elsarticle %package on \href{http://www.ctan.org/tex-%archive/macros/latex/contrib/elsarticle}{CTAN}.}

%% Group authors per affiliation:
%\author{Elsevier\fnref{myfootnote}}
%\address{Radarweg 29, Amsterdam}
%\fntext[myfootnote]{Since 1880.}

%% or include affiliations in footnotes:
%\author[mymainaddress,mysecondaryaddress]{Elsevier Inc}
%\ead[url]{www.elsevier.com}

\author[aff1]{Manash Dey \fnref{present1}}
\fntext[present1]{Present affiliation: Department of Physics, Maibang Degree College, India, 788831.}

\ead{manashdey@gauhati.ac.in; manashdey1272@gmail.com}

\address[aff1]{Department of Physics, Gauhati University, India, 781014.}

\begin{abstract}
We posit an elegant modular $A_4$ framework for Dirac neutrinos that does not rely on the seesaw mechanism in a non SUSY setting. Our construction ensures purely Dirac neutrinos with a minimal scalar sector, naturally generating neutrino masses without requiring unnaturally tiny couplings. The model demonstrates its predictive power by simultaneously reproducing the charged lepton mass hierarchy and predicting neutrino mixing angles and mass squared differences consistent with their global best fit values. It further predicts a sum of neutrino masses consistent with the current cosmological bound, while predicting maximal Dirac CP violation. This construction establishes an alternative paradigm for the origin of lepton masses and mixing, setting it apart from conventional discrete flavour symmetry and seesaw based approaches.
\end{abstract}

%\keywords{Suggested keywords}%Use showkeys class option if keyword
                              %display desired
\maketitle

%\tableofcontents

The discovery of neutrino oscillations has shown that neutrinos have mass and that neutrino flavours mix, providing the first clear evidence of physics beyond the Standard Model (SM). While the neutrino mixing angles and mass squared differences have been measured with good precision, the origin of neutrino masses and the observed mixing patterns remain unknown.

Discrete flavour symmetries\,\cite{Ishimori:2010au, Altarelli:2010gt} have long been explored to address these questions, but such models typically require multiple flavon fields with carefully aligned vacuum expectation values\,(vevs), thereby adding complexity. Many of them rely on the seesaw mechanism\,\cite{Yoshimura:1978ex}, which usually assumes Majorana neutrinos. However, the Majorana nature remains experimentally unconfirmed due to the non observation of neutrinoless double beta decay\,($0\nu\beta\beta$)\,\cite{Schechter:1981bd}. This motivates us to explore Dirac neutrino frameworks. 

	In such scenarios, the smallness of neutrino masses can be achieved via the Dirac seesaw mechanism\,\cite{Lindner:2001hr}, analogous to the Majorana case. While well motivated, we argue that small neutrino masses can also be generated in a Dirac framework without invoking any seesaw mechanism. However, in such non seesaw scenarios, the renormalisable tree level term is generally allowed, and reproducing sub eV neutrino masses then requires the tree level Yukawa couplings to be unnaturally small. Therefore, any realistic non seesaw Dirac model must first forbid the tree level Dirac Yukawa and then generate neutrino masses through a suitably suppressed higher dimensional operator.

	Interestingly, in one of our earlier works \cite{Dey:2024ctx}, we constructed a Dirac model without a seesaw mechanism, in which the tree level term was forbidden by discrete symmetries. In that setup, tiny neutrino masses emerged naturally from the interplay of flavon vevs and the cutoff scale, thereby avoiding both the seesaw mechanism and unnaturally small Yukawa couplings. While successful, that construction required an extended scalar sector, reducing its minimality.
	
	 A natural way to overcome this limitation is to leverage the power of modular symmetry\,\cite{Feruglio:2017spp, narayanan2019modular, Ding:2023htn, Kobayashi:2023zzc}, where both matter fields and Yukawa couplings transform under a finite modular subgroup $\Gamma_N$, with Yukawa couplings represented by modular forms. This framework substantially reduces the need for flavon fields and provides a minimal and elegant approach to understanding lepton masses and mixing. Although most studies implement modular symmetry in a supersymmetric (SUSY) context, low energy SUSY remains experimentally elusive, motivating the exploration of non SUSY modular frameworks. Recently `Qu' and `Ding' showed that modular symmetries can be consistently formulated in a non holomorphic framework\,\cite{Qu:2024rns}, enabling their application in non SUSY models\,\cite{Nomura:2024vzw, Zhang:2025dsa, Qu:2025ddz, Nomura:2025raf, Priya:2025wdm, Abbas:2025nlv, Li:2024svh, Nomura:2024nwh, Nomura:2024atp}. This non holomorphic modular approach therefore offers a predictive framework for flavour model building, while simultaneously reducing the reliance on an extended scalar sector.

	In this work, we adopt a novel approach by presenting a non SUSY modular framework, based on the finite modular subgroup $\Gamma_3 \simeq A_4$, in which neutrinos are purely Dirac particles, independent of the seesaw mechanism. Although seesaw realizations have been the dominant route in modular symmetry models, our framework demonstrates that neutrino masses can arise naturally without invoking a seesaw and without requiring unnaturally suppressed couplings.

	We extend the SM by incorporating the modular $A_4$ flavour symmetry\,\cite{Altarelli:2005yx, Mishra:2022egy, Kashav:2021zir, Kashav:2022kpk, Singh:2024imk, Kashav:2024lkr, Behera:2020lpd, Behera:2020sfe, CentellesChulia:2023osj, Kumar:2023moh, Devi:2023vpe}. The field content relevant for the lepton sector is as follows: the lepton doublets $\overline{L}_l$ transform as a triplet under $A_4$ with weight $k=1$, the right handed neutrinos $\nu_R$ transform as a triplet under $A_4$ with weight $-1$, and the right handed charged leptons $(e_R, \mu_R, \tau_R)$ transform as $1, 1'', 1'$ under $A_4$ with weights $5$, $3$, and $2$, respectively. The Higgs doublet $H$ is an $A_4$ singlet with weight $0$, and the singlet scalars $\rho, \sigma, \eta$ are all $A_4$ singlets with weights $+2, -1,$ and $+1$, respectively. In our construction, only modular forms of weights $+2$ and $+4$ are used, ensuring that Majorana mass terms for $\nu_R$ are absent and neutrinos remain Dirac. However, a modular form of weight $-2$ could, in principle, reintroduce Majorana operators. While it is common in modular models to assume that certain modular weights do not appear in a theory. In our model, this would correspond to weight $-2$. However, such an assumption is not fully justified. To ensure all Majorana terms are forbidden, we instead impose an additional $Z_{10}$ symmetry\,\cite{Dey:2023rht, Dey:2024ctx, Dey:2023bfa, Dey:2025sqe}. Under this symmetry, only $\overline{L}_l$, the right handed charged leptons, and $\nu_R$ transform non trivially, with charges $6$, $4$, and $4$, respectively. These assignments consistently forbid all Majorana operators, maintaining a purely Dirac neutrino framework.
		
	With the field content and symmetry assignments outlined above, the Yukawa Lagrangian of our model, consistent with both $A_4$ modular symmetry and the $Z_{10}$ symmetry, is constructed in the following way
	 
\begin{eqnarray}
-\mathcal{L} &=&  \frac{\alpha_e \textbf{Y}^{(2)}_3}{\Lambda^4} \overline{L}_l H e_R\, \sigma^4 + \frac{\beta_{\mu} \textbf{Y}^{(2)}_3}{\Lambda^2} \overline{L}_l H \mu_R\, \sigma^2 + \frac{\gamma_{\tau} \textbf{Y}^{(2)}_3}{\Lambda} \nonumber\\&& \overline{L}_l H \tau_R\, \sigma + \frac{y_{1,2} \textbf{Y}^{(2)}_3}{\Lambda} (\overline{L}_l\, \nu_R)_{s,a} \tilde{H}\, \rho + \frac{y_{3,4} \textbf{Y}^{(4)}_3}{\Lambda^3} (\overline{L}_l\nonumber\\&& \nu_R)_{s,a} \tilde{H}\, \rho\, \eta^2 + \frac{y_5}{\Lambda^3} \overline{L}_l \nu_R \tilde{H}\, \rho \,\sigma^2 + \frac{y_6 \textbf{Y}^{(4)}_{1^{'}}}{\Lambda^3} \overline{L}_l \nu_R \tilde{H} \nonumber\\&& \, \rho\, \eta^2 + h.c.
\end{eqnarray}
Following the decomposition of the $A_4$ representations and using the modular forms $\mathbf{Y}^{(2)}_3$, $\mathbf{Y}^{(4)}_{1'}$, and $\mathbf{Y}^{(4)}_3$ \,\cite{Qu:2024rns}, we can construct the fermion mass matrices. 

The charged lepton mass matrix ($M_l$) then takes the form shown below

\begin{eqnarray}
M_{l} &=& v_h
\begin{pmatrix}
  \alpha_e Y_1 (\frac{v_{\sigma}}{\Lambda})^4 & \beta_\mu Y_2 (\frac{v_{\sigma}}{\Lambda})^2  &  \gamma_\tau Y_3 (\frac{v_{\sigma}}{\Lambda}) \\
  \alpha_e Y_3 (\frac{v_{\sigma}}{\Lambda})^4    &  \beta_\mu Y_1 (\frac{v_{\sigma}}{\Lambda})^2 & \gamma_\tau Y_2 (\frac{v_{\sigma}}{\Lambda})  \\
  \alpha_e Y_2 (\frac{v_{\sigma}}{\Lambda})^4    &  \beta_\mu Y_3 (\frac{v_{\sigma}}{\Lambda})^2 & \gamma_\tau Y_1 (\frac{v_{\sigma}}{\Lambda})
\end{pmatrix}.
\end{eqnarray}
The neutrino mass matrix\,($M_\nu$) of the model is given by
\begin{eqnarray}
M_{\nu} &=& 
\begin{pmatrix}
  a & b  & c\\
  g & d  &  e\\
  h & p  &  f
\end{pmatrix},
\end{eqnarray}
where the matrix elements are explicitly shown below
\setlength{\jot}{2pt}
\setlength{\arraycolsep}{2pt}
\begin{eqnarray}
a &=& \frac{2\,y_1 v_h \, v_\rho Y_1}{3 \Lambda} +  \frac{2\, y_3\, v_h \, v_\rho v^2_\eta (Y^2_1 - Y_2 Y_3)}{3 \Lambda^3} + \frac{ y_5\, v_h\, v_\rho v^2_\sigma}{\Lambda^3} ,\nonumber\\
b &=& -\frac{v_h \, v_\rho Y_3}{\Lambda}(\frac{y_1}{3}- \frac{y_2}{2}) -\frac{v_h \, v_\rho v^2_\eta (Y^2_2 - Y_1 Y_3)}{\Lambda^3}(\frac{y_3}{3}- \frac{y_4}{2}),\nonumber\\
c &=& -\frac{v_h \, v_\rho Y_2}{\Lambda}(\frac{y_1}{3}+ \frac{y_2}{2})-\frac{v_h \, v_\rho v^2_\eta (Y^2_3 - Y_1 Y_2)}{\Lambda^3}(\frac{y_3}{3}+ \frac{y_4}{2})\nonumber\\&& \quad + \frac{y_6\,v_h\, v_\rho v^2_\eta}{\Lambda^3}(Y^2_3 + 2 Y_1 Y_2),\nonumber\\
d &=& \frac{2\, y_1\, v_h \, v_\rho Y_2}{3 \Lambda} + \frac{2\,y_3\, v_h \, v_\rho v^2_\eta (Y^2_3 - Y_1 Y_2)}{3 \Lambda^3}+ \frac{y_6\,v_h\, v_\rho v^2_\eta}{\Lambda^3}\nonumber\\&& \quad(Y^2_3 + 2 Y_1 Y_2),\nonumber\\
e &=& -\frac{v_h \, v_\rho Y_1}{\Lambda}(\frac{y_1}{3}- \frac{y_2}{2})-\frac{v_h \, v_\rho v^2_\eta (Y^2_1 - Y_2 Y_3)}{\Lambda^3}(\frac{y_3}{3}- \frac{y_4}{2})\nonumber\\&& \quad + \frac{ y_5\, v_h\, v_\rho v^2_\sigma}{\Lambda^3},\nonumber\\
f &=& \frac{2\,y_1\,v_h \, v_\rho Y_3}{3 \Lambda} + \frac{2\, y_3\,v_h \, v_\rho v^2_\eta (Y^2_2 - Y_1 Y_3)}{3 \Lambda^3},\nonumber\\
g &=& -\frac{v_h \, v_\rho Y_3}{\Lambda}(\frac{y_1}{3}+ \frac{y_2}{2})-\frac{v_h \, v_\rho v^2_\eta (Y^2_2 - Y_1 Y_3)}{\Lambda^3}(\frac{y_3}{3}+ \frac{y_4}{2}),\nonumber\\
h &=& -\frac{v_h \, v_\rho Y_2}{\Lambda}(\frac{y_1}{3}- \frac{y_2}{2})-\frac{v_h \, v_\rho v^2_\eta (Y^2_3 - Y_1 Y_2)}{\Lambda^3}(\frac{y_3}{3}- \frac{y_4}{2}) \nonumber\\&& \quad+ \frac{ y_6\,v_h\, v_\rho v^2_\eta}{\Lambda^3}(Y^2_3 + 2 Y_1 Y_2),\nonumber\\
p &=& -\frac{v_h \, v_\rho Y_1}{\Lambda}(\frac{y_1}{3}+ \frac{y_2}{2})-\frac{v_h \, v_\rho v^2_\eta (Y^2_1 - Y_2 Y_3)}{\Lambda^3}(\frac{y_3}{3}+ \frac{y_4}{2})\nonumber\\&& \quad + \frac{y_5\,v_h\, v_\rho v^2_\sigma}{\Lambda^3}.
\end{eqnarray}
Here, in the $M_l$ and $M_\nu$, $Y_1$, $Y_2$, and $Y_3$ are the components of the modular triplet $\textbf{Y}_3^{(2)}$\,\cite{Qu:2024rns} and are functions of the modulus $\tau$. The quantities $v_h$, $v_\rho$, $v_\sigma$, and $v_\eta$ are the scalar vevs, $\Lambda$ is the cutoff scale, $\alpha_e$, $\beta_\mu$, and $\gamma_\tau$ are real parameters, and $y_i$ ($i=1\dots6$) are complex. In the absence of a seesaw mechanism, the modular forms, in conjunction with the scalar vevs and the cutoff scale, naturally account for the observed lepton mass scales.

	Having established the structure of the mass matrices and the relevant parameters, we now proceed to the numerical analysis of the model.

	The charged lepton mass matrix $M_l$ is diagonalized by a unitary matrix $U_l$ as $U_l^\dagger M_l M_l^\dagger U_l = \text{diag}(m_e^2, m_\mu^2, m_\tau^2)$, where $m_e$, $m_\mu$, and $m_\tau$ are the charged lepton masses. The observed hierarchy among the charged lepton masses is realized through the suppression factors of order $v_\sigma/\Lambda$ present in $M_l$. 

	For the neutrino sector, the Dirac mass matrix $M_\nu$ is diagonalized as $U_\nu^\dagger M_\nu^\dagger M_\nu U_\nu = \text{diag}(m_1^2, m_2^2, m_3^2)$ with $m_1, m_2, m_3$ denoting the neutrino mass eigenvalues. The lepton mixing matrix is then obtained as $U = U_l^\dagger U_\nu$, from which the mixing angles $(\theta_{12}, \theta_{13}, \theta_{23})$ and the Dirac CP violating phase $\delta$ are extracted using
\begin{align}
\sin^2 \theta_{13} = |U_{13}|^2, \qquad
\sin^2 \theta_{12} = \frac{|U_{12}|^2}{1 - |U_{13}|^2}, \nonumber \\
\sin^2 \theta_{23} = \frac{|U_{23}|^2}{1 - |U_{13}|^2}, \qquad
\sin\delta = \frac{J_{\text{CP}}}{c_{12} s_{12} c_{23} s_{23} c_{13}^2 s_{13}},
\end{align}  
where, $J_{\text{CP}} = \operatorname{Im}\left[ U_{11} U_{22} U_{12}^* U_{21}^* \right]$ is the Jarlskog invariant, and $c_{ij} = \cos\theta_{ij}$, $s_{ij} = \sin\theta_{ij}$. For the numerical analysis, we fix $v_h = 246~\text{GeV}$, $\Lambda = 10^{14}~\text{GeV}$, $v_\sigma = 10^{12}~\text{GeV}$, $\beta_\mu/\alpha_e = 0.0203$, and $\gamma_\tau/\alpha_e = 0.003$, and perform a numerical scan over the remaining parameters, including the modulus $\tau$ restricted to its fundamental domain\,\cite{narayanan2019modular}. The objective is to identify regions of parameter space that yield a hierarchical charged lepton spectrum while simultaneously reproducing the mixing angles  and the Dirac CP violating phase, and the mass squared differences $(\Delta m_{21}^2, \Delta m_{31}^2)$ at their global\,(NuFIT 6.0) best fit values\,\cite{Esteban:2024eli}. In addition, the sum of neutrino masses $\sum m_i$ is required to remain consistent with cosmological bound\,\cite{Planck:2018vyg}. Although other regions of the parameter space can reproduce the observables within their $3\sigma$ ranges, our focus here is on identifying the regions of parameter space that are in excellent agreement with the global best fit values, as these correspond to the most probable configuration. The scan is carried out for both normal hierarchy (NH) and inverted hierarchy (IH).

	After performing the scan, we find that the model remains viable for both NH and IH. In each case, the corresponding ranges of the numerical parameters consistent with our criteria are summarized in Table~\ref{tab:NH_IH}.
\begin{table}[h!]
\centering\setlength{\tabcolsep}{1.99pt}
\renewcommand{\arraystretch}{1.4}
\begin{tabular}{lc c c c}
\hline
 & \multicolumn{2}{c}{\textbf{NH}} & \multicolumn{2}{c}{\textbf{IH}} \\
\cline{2-3} \cline{4-5}
\textbf{Parameter} & \textbf{Min} & \textbf{Max} & \textbf{Min} & \textbf{Max} \\
\hline
$\theta_{12}$ & \multicolumn{2}{c}{$33.60^\circ$} & \multicolumn{2}{c}{$33.68^\circ$} \\
$\theta_{13}$ & \multicolumn{2}{c}{$8.50^\circ$}  & \multicolumn{2}{c}{$8.58^\circ$} \\
$\theta_{23}$ & \multicolumn{2}{c}{$48.5^\circ$}  & \multicolumn{2}{c}{$48.6^\circ$} \\
$\delta$      & \multicolumn{2}{c}{$270^\circ$}   & \multicolumn{2}{c}{$270^\circ$} \\
$\Delta m_{21}^2/10^{-5}$ & \multicolumn{2}{c}{$7.50~\text{eV}^2$} & \multicolumn{2}{c}{$7.50~\text{eV}^2$} \\
$\Delta m_{31}^2/10^{-3}$ & \multicolumn{2}{c}{$2.534 ~\text{eV}^2$} & \multicolumn{2}{c}{$-2.512~\text{eV}^2$} \\
$Re[\tau]$& $-0.424$ & $0.331$ & $-0.457$ & $0.490$ \\
$Im[\tau]$& $1.001$ & $1.521$ & $1.016$ & $1.532$ \\
$m_e$ & $0.060 ~\text{MeV}$ & $0.541 ~\text{MeV}$ & $0.337 ~\text{MeV}$ & $0.586 ~\text{MeV}$  \\
$m_{\mu}$ & $104 ~\text{MeV}$ & $119 ~\text{MeV}$ & $103 ~\text{MeV}$ &  $123 ~\text{MeV}$  \\
$m_{\tau}$ & $1510 ~\text{MeV}$ & $1892 ~\text{MeV}$ &  $1508 ~\text{MeV}$  &  $1830 ~\text{MeV}$  \\
$|v_{\eta}|/10^{12}$ & $8.752~\text{GeV}$ & $9.350~\text{GeV}$ & $8.147~\text{GeV}$ & $16~\text{GeV}$ \\
$v_{\rho}$ & $315 ~\text{GeV}$ & $992 ~\text{GeV}$ & $270~\text{GeV}$ & $996~\text{GeV}$ \\
\hline
$Re[y_1]$ & $-0.006$ & $0.016$ & $-0.053$ & $0.039$ \\
$Im[y_1]$ & $-0.032$ & $0.021$ & $-0.061$ & $0.051$ \\
$Re[y_2]$ & $-0.021$ & $0.030$ & $-0.033$ & $0.054$ \\
$Im[y_2]$ & $-0.008$ & $0.061$ & $-0.040$ & $0.072$ \\
$Re[y_3]$ & $-4.866$ & $4.946$ & $-3.735$ & $4.816$ \\
$Im[y_3]$ & $-4.977$ & $4.874$ & $-4.345$ & $4.591$ \\
$Re[y_4]$ & $-4.377$ & $1.705$ & $-4.846$ & $4.864$ \\
$Im[y_4]$ & $-3.127$ & $4.994$ & $-4.398$ & $3.996$ \\
$Re[y_5]$ & $-7.580$ & $5.288$ & $-4.804$ & $4.829$ \\
$Im[y_5]$ & $-4.871$ & $5.233$ & $-4.881$ & $4.608$ \\
$Re[y_6]$ & $-1.889$ & $3.754$ & $-2.046$ & $3.032$ \\
$Im[y_6]$ & $-3.820$ & $2.343$ & $-1.101$ & $2.745$ \\
\hline
\end{tabular}
\caption{Predicted values of observables and the allowed parameter ranges obtained from the numerical scan for both NH and IH.}
\label{tab:NH_IH}
\end{table} 
	From Table~\ref{tab:NH_IH}, it is evident that, for both NH and IH, the allowed parameter space is tightly constrained by the best fit requirement. The modulus $\tau$ lies within a narrow band of the fundamental domain, with $Im[\tau]$ clustering around $1.0-1.5$; for these values of $\tau$, the neutrino mixing angles and mass squared differences are reproduced in excellent agreement with their global best fit values. In the NH case, while the CP phase $\delta$ does not exactly coincide with the best fit point, the prediction $\delta \approx 270^\circ$ remains consistent with the NuFIT 6.0 global analysis\,\cite{Esteban:2024eli}, favouring maximal CP violation. In the charged lepton sector, the observed mass hierarchy is realized, with $m_e \simeq 0.5~\text{MeV}$, $m_\mu \simeq 105~\text{MeV}$, and $m_\tau \simeq 1777~\text{MeV}$. The framework also predicts a sharply determined sum of neutrino masses, $\Sigma m_i \simeq 0.10$ eV (NH) and $\Sigma m_i \simeq 0.115$ eV (IH), which is consistent with current cosmological upper bound\,\cite{Planck:2018vyg}. As noted previously, owing to the rich structure of the model, a broader scan of the parameter space can reproduce other regions of the $3\sigma$ ranges of the observables. Consequently, even if future global fits deviate from the current best fit values, the model can be tuned within this extended parameter space to maintain consistency, thereby preserving its predictive power and phenomenological viability. In addition, the scalar vevs are driven into distinct ranges: $v_\eta \sim \mathcal{O}(10^{12})~\text{GeV}$ and $v_\rho$ in the electroweak – TeV window ($300$–$1000$ GeV). The parameters $y_i$ are found to take reasonable values compatible with the charged lepton mass hierarchy and the neutrino sector observables.

\begin{figure}[h!]
    \centering
    \begin{minipage}[b]{0.23\textwidth}
        \centering
        \includegraphics[width=\textwidth]{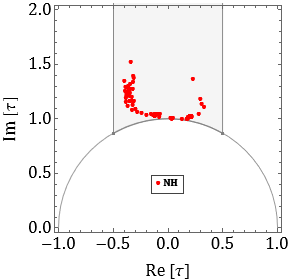}
    \end{minipage}
    \hfill
    \begin{minipage}[b]{0.23\textwidth}
        \centering
        \includegraphics[width=\textwidth]{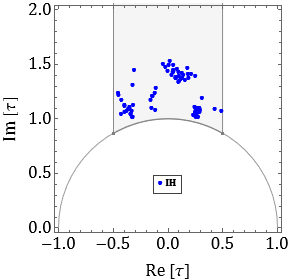}
    \end{minipage}
    \caption{The plots highlight the allowed regions of the modulus $\tau$ in the fundamental domain for NH and IH.}
    \label{fig:1}
\end{figure}

\begin{figure}[h!]
    \centering
    \begin{minipage}[b]{0.237\textwidth}
        \centering
        \includegraphics[width=\textwidth]{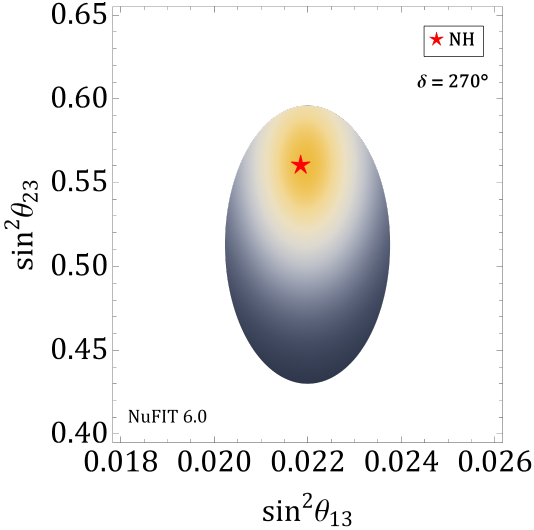}
    \end{minipage}
    \hfill
    \begin{minipage}[b]{0.237\textwidth}
        \centering
        \includegraphics[width=\textwidth]{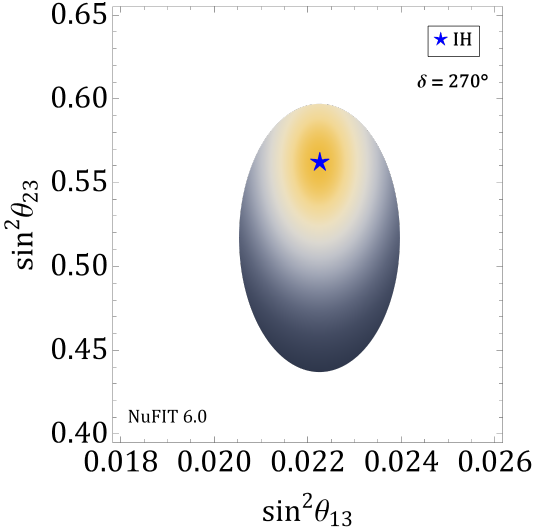}
    \end{minipage}

    \vspace{0.4cm}
    \begin{minipage}[b]{0.237\textwidth}
        \centering
        \includegraphics[width=\textwidth]{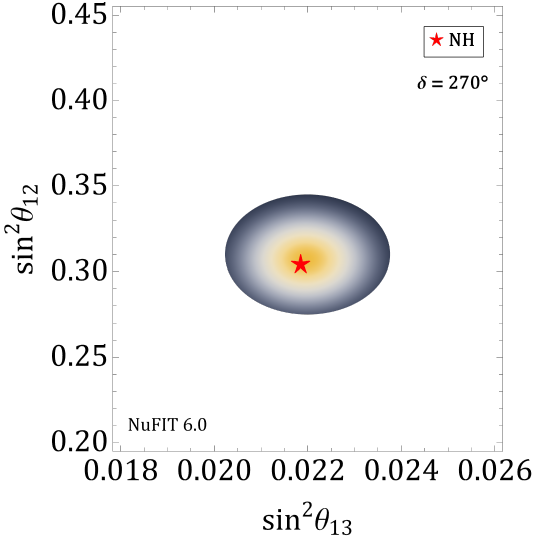}
    \end{minipage}
    \hfill
    \begin{minipage}[b]{0.238\textwidth}
        \centering
        \includegraphics[width=\textwidth]{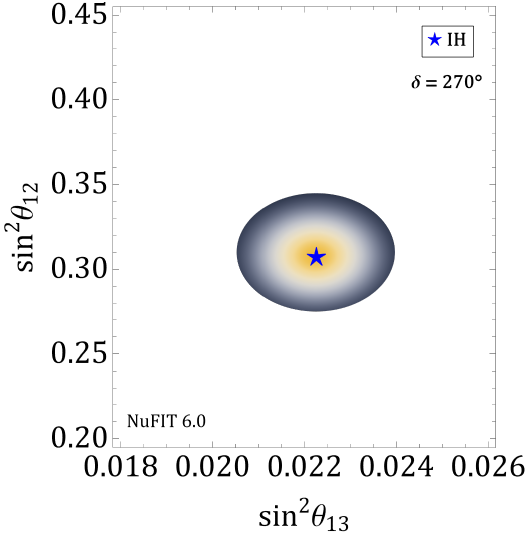}
    \end{minipage}

    \vspace{0.4cm}
    \begin{minipage}[b]{0.217\textwidth}
        \centering
        \includegraphics[width=\textwidth]{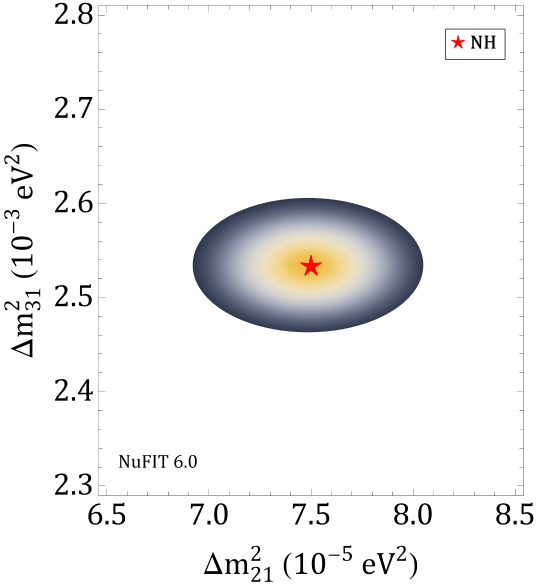}
    \end{minipage}
    \hfill
    \begin{minipage}[b]{0.226\textwidth}
        \centering
        \includegraphics[width=\textwidth]{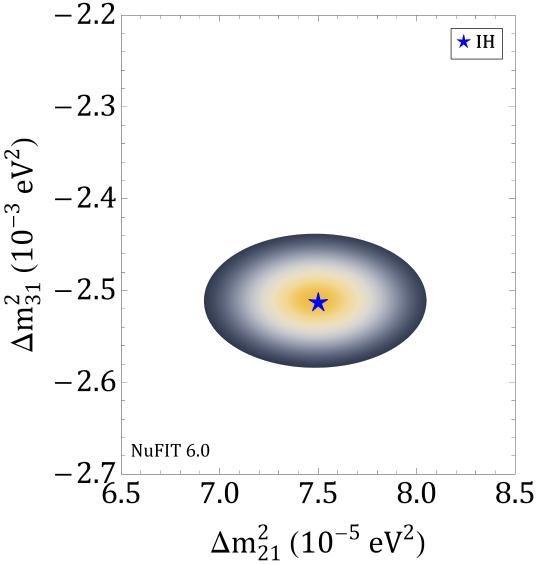}
    \end{minipage}
    
    \vspace{0.4cm}
    
    \begin{minipage}[b]{0.224\textwidth}
        \centering
        \includegraphics[width=\textwidth]{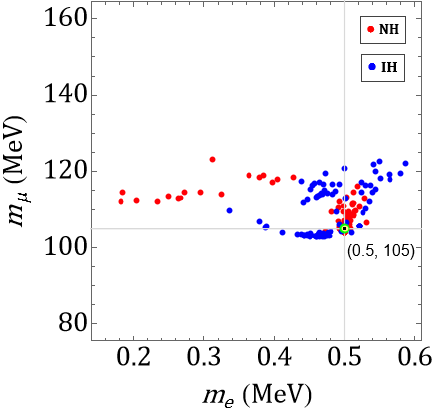}
    \end{minipage}
    \hfill
    \begin{minipage}[b]{0.224\textwidth}
        \centering
        \includegraphics[width=\textwidth]{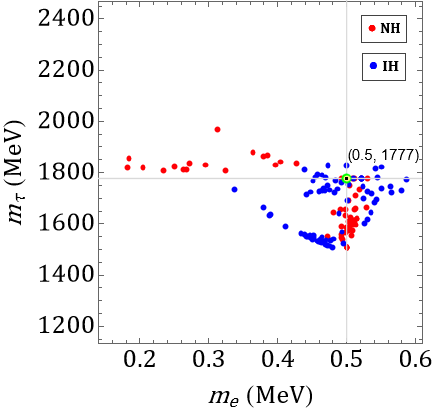}
    \end{minipage}
    
    \vspace{0.4cm}
    \caption{The plots highlight the allowed regions of lepton sector observables for NH and IH. For the neutrino mixing angles and mass squared differences\,(top three rows), the $3\sigma$ ranges are shown as density regions: yellow marks the best fit vicinity, fading through whitish tones to darker gray with increasing deviation from the global best fit values. The red (NH) and blue (IH) stars denote the model predictions which are in excellent agreement with the global best fit values. For the charged lepton masses\,(bottom row), the points marked in brackets correspond to the measured experimental values, demonstrating that the model reproduces the observed charged lepton mass hierarchy accurately.}
    \label{fig:2}
\end{figure}

\begin{figure}[h!]
    \centering
    
     \vspace{0.14cm}
     
    \begin{minipage}[b]{0.238\textwidth}
        \centering
        \includegraphics[width=\textwidth]{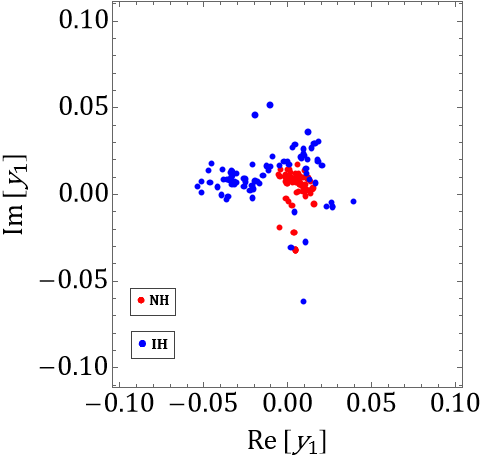}
    \end{minipage}
    \hfill
    \begin{minipage}[b]{0.238\textwidth}
        \centering
        \includegraphics[width=\textwidth]{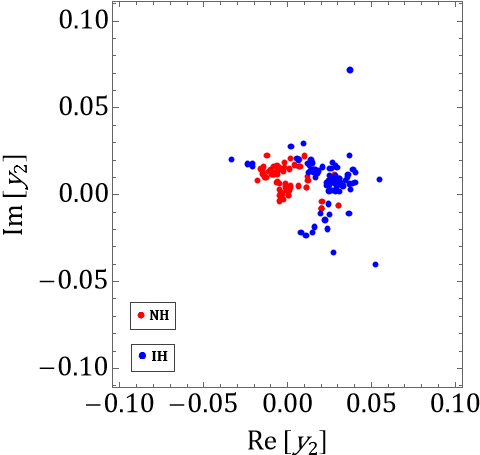}
    \end{minipage}

    \vspace{0.58cm}
    \begin{minipage}[b]{0.228\textwidth}
        \centering
        \includegraphics[width=\textwidth]{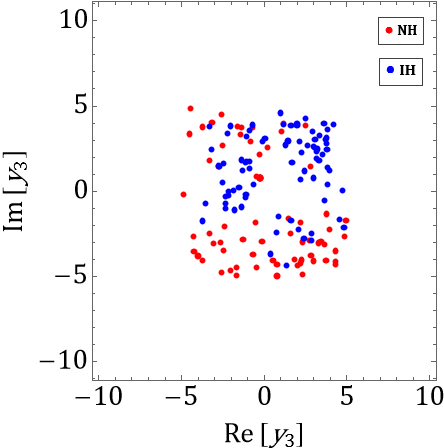}
    \end{minipage}
    \hfill
    \begin{minipage}[b]{0.228\textwidth}
        \centering
        \includegraphics[width=\textwidth]{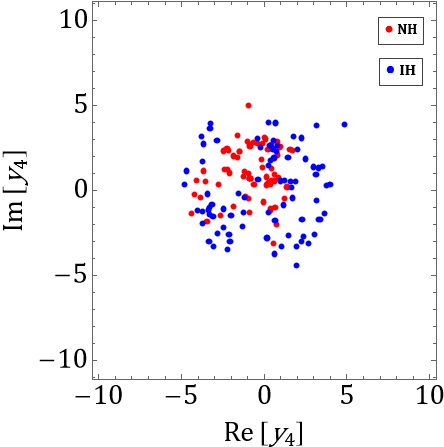}
    \end{minipage}

    \vspace{0.5cm} 
    \begin{minipage}[b]{0.228\textwidth}
        \centering
        \includegraphics[width=\textwidth]{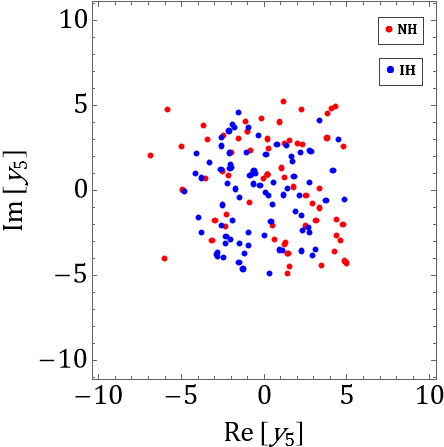}
    \end{minipage}
    \hfill
    \begin{minipage}[b]{0.228\textwidth}
        \centering
        \includegraphics[width=\textwidth]{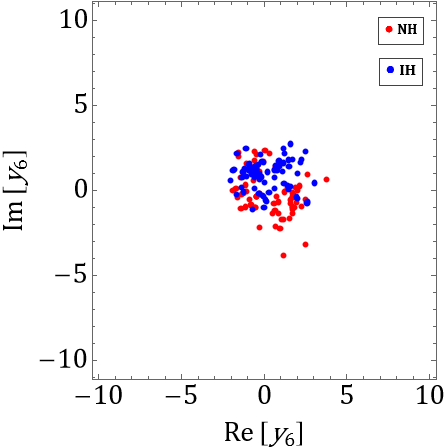}
    \end{minipage}

    \vspace{0.49cm} 
    \begin{minipage}[b]{0.24\textwidth}
        \centering
        \includegraphics[width=\textwidth]{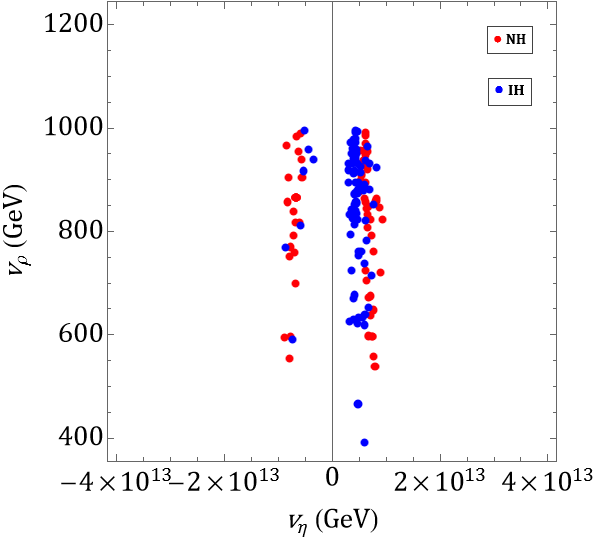}
    \end{minipage}
    \caption{Plots showing the allowed parameter regions for the couplings $y_i$ and the scalar vevs $v_\eta$ and $v_\rho$ for NH and IH.}
    \label{fig:3}
\end{figure}

	To provide a graphical understanding of the predicted parameter space, we highlight representative plots for both NH and IH. In Fig.~\ref{fig:1}, we show $Re[\tau]$ versus $Im[\tau]$. In Fig.~\ref{fig:2}, we present scatter plots of the mixing angles, mass squared differences, and charged lepton masses. In Fig.~\ref{fig:3}, we display scatter plots of the remaining model parameters.

	In sum, we present a novel, seesaw evaded Dirac neutrino framework based on modular $A_4$ symmetry. In this framework, sub eV neutrino masses arise naturally from the interplay of modular forms, three singlet scalar vevs, and the cutoff scale, without requiring unnaturally small couplings. The model uniquely predicts all neutrino mixing angles and mass squared differences consistent with the global best fit values. In addition, it yields a sum of neutrino masses compatible with cosmological bounds and predicts maximal Dirac CP violation. Owing to its economy of fields and predictive character, this framework offers a fresh perspective on the origin of neutrino masses and mixing beyond traditional discrete flavour symmetry and seesaw based approaches.
\section*{\noindent\textit{Acknowledgements}.}
MD is thankful to Subhankar Roy, Department of Physics, Gauhati University, for his kind support and encouragement.
MD would also like to thank Amol Dighe (TIFR, Mumbai) and S. Uma Sankar (IIT Bombay) for valuable discussions and comments on this work during the IIFC-vP Neutrino School at NISER, Bhubaneswar, 2025. MD also acknowledges the Council of Scientific and Industrial Research (CSIR), Government of India, New Delhi, for financial support through a NET Senior Research Fellowship, vide grant No. 09/0059(15346)/2022-EMR-I, under which this work was carried out.
\biboptions{sort&compress}
\bibliography{references} % your cleaned .bib file
\end{document}